\documentclass{raa}      
\usepackage{graphicx,times}
\usepackage{natbib}
\usepackage{amssymb,amsmath}
\bibpunct{(}{)}{;}{a}{}{,}

\newcommand{\full}{${\tt full\_1}$}
\newcommand{\vmax}{${V_{\rm max}}$}

\newcommand{\mpch}{{h^{-1}\rm Mpc}}

\newcommand{\kms}{{\rm km}\,\mathrm{s}^{-1}}

\usepackage[a4paper=true,dvipdfm=true,pagebackref=true]{hyperref}
\hypersetup{pdftitle = The title of my PDF, pdfauthor = My name, pdfsubject= The subject, pdfkeywords = keyword1 keyword2 keyword3} 
\hypersetup{colorlinks = true, linkcolor = green, anchorcolor = red, citecolor = blue, filecolor = red, pagecolor = red, urlcolor = red}

\begin{document}

   \title{Toward accurate measurement of property-dependent galaxy clustering}
   \subtitle{I. Comparison of the \vmax\ method and the ``shuffled'' method}

 \volnopage{ {\bf 20XX} Vol.\ {\bf X} No. {\bf XX}, 000--000}
   \setcounter{page}{1}

   \author{Lei Yang
      \inst{1}, Yipeng Jing
      \inst{1,2}, Zhigang Li,
      \inst{3}, Xiaohu Yang
      \inst{1,2}
   }

   \institute{Department of Astronomy, School of Physics and Astronomy, Shanghai Jiao Tong University, Shanghai, 200240, China; {\it ypjing@sjtu.edu.cn}\\
        \and
             Tsung-Dao Lee Institute, and Shanghai Key Laboratory for Particle Physics and Cosmology, Shanghai Jiao Tong University, Shanghai, 200240, China\\
        \and
             College of Physics and Electronic Engineering, Nanyang Normal University, Nanyang, Henan, 473061, China\\
\vs \no
   {\small Received 2019-11-15; accepted 2019-12-11}
}

\abstract{Galaxy clustering provides insightful clues to our understanding of galaxy formation and 
evolution, as well as the universe. The redshift assignment for the random sample is one of the 
key steps to measure the galaxy clustering accurately. In this paper, by virtue of the mock galaxy 
catalogs, we investigate the effect of two redshift assignment methods on the measurement of 
galaxy two-point correlation functions (hereafter 2PCFs), the \vmax\ method and the ``shuffled'' method. 
We found that the shuffled method significantly underestimates both of the projected 
2PCFs and the two-dimensional 2PCFs in redshift space. While the \vmax\ method 
does not show any notable bias on the 2PCFs for volume-limited samples. 
For flux-limited samples, the bias produced by the \vmax\ method is less than half of 
the shuffled method on large scales. Therefore, we strongly recommend the \vmax\ method 
to assign redshifts to random samples in the future galaxy clustering analysis.
}

   \authorrunning{Yang et al. }            
   \titlerunning{Toward accurate measurement of property-dependent galaxy clustering}  
   \maketitle

%
\section{Introduction}           
\label{sect:intro}
Observed galaxy distribution encodes a wealth of information on the
formation and evolution of galaxies, dark matter halos, and the 
large-scale structure of the universe. In the past two decades, with
the successes of completed and ongoing wide-field surveys such as the
Two Degree Field Galaxy Redshift Survey (2dFGRS;
\citealt{2003astro.ph..6581C}), the Sloan Digital Sky Survey (SDSS;
\citealt{2000AJ....120.1579Y}), the Baryon Oscillation Spectroscopic
Survey (BOSS; \citealt{2011AJ....142...72E}), the VIMOS Public
Extragalactic Redshift Survey (VIPERS; \citealt{2012PASP..124.1232G}),
and the Dark Energy Spectroscopic Instrument (DESI;
\citealt{2013arXiv1308.0847L,2016arXiv161100036D,2016arXiv161100037D}),
we are able to map the three-dimensional distribution of over a
million galaxies with well-measured spectroscopic redshifts. These
observed galaxies exhibit a variety of physical properties (e.g.,
luminosity, color, stellar mass, morphology, spectral type) as well as 
notable environment-dependent features
\citep{1997ApJ...490..577D,2003ApJ...594..186B,
  2003MNRAS.346..601G}. Consequently one primary goal of observational
cosmology is to utilize an efficient and reliable technique to
optimally extract information from these samples, in order to
interpret these property-dependent distributions and gain some
cosmological insights.

The galaxy two-point correlation function is one of
the most powerful and fundamental tools to characterize the spatial
distribution of galaxies \citep{1980lssu.book.....P}.  On small
scales, apart from  the galaxy peculiar velocities
\citep{1972MNRAS.156P...1J, 2003MNRAS.346...78H,2013A&A...557A..54D}, 
the 2PCF is shaped by the complex baryonic physics involved in
galaxy formation in dark matter halos, offering unique checks for
empirical galaxy-halo connection models, e.g., the halo occupation
distribution model (HOD;
\citealt{1998ApJ...494....1J,2002ApJ...564...15J,
 2000MNRAS.318.1144P,2002ApJ...575..587B,2005ApJ...633..791Z,
 2015MNRAS.453.4368G,2018MNRAS.481.5470X}),
the conditional luminosity function technique (CLF;
\citealt{2003MNRAS.339.1057Y, 2004MNRAS.350.1153Y,2005MNRAS.358..217Y,
  2005MNRAS.357..608Y,2008ApJ...676..248Y,
  2012ApJ...752...41Y,2018ApJ...860...30Y,2004MNRAS.353..189V,2007MNRAS.376..841V}),
and the subhalo abundance matching method
(SHAM;\citealt{2004ApJ...609...35K,
  2006ApJ...647..201C,2006MNRAS.371.1173V,2010MNRAS.404.1111G,
  2012MNRAS.423.3458S, 2014MNRAS.437.3228G, 2016MNRAS.460.3100C}). 
On large scales, the anisotropy imprinted in the redshift-space
clustering, arising from the gravity-driven coherent motion of matter,
is widely used to measure the growth rate of the cosmic structure, to
distinguish dark energy models and to constrain the
cosmological parameters \citep{1987MNRAS.227....1K,
  2001Natur.410..169P, 2004PhRvD..69j3501T, 2006JCAP...10..014S,
  2008Natur.451..541G, 2010MNRAS.401.2148P,2011MNRAS.415.2876B,
  2012MNRAS.426.2719R, 2013PhR...530...87W, 2014MNRAS.437.1109R,
  2016ApJ...833..287L, 2018ApJ...861..137S,
  2018MNRAS.481.3160W}. Therefore, to accurately measure the 2PCF is a
critical step for probing the galaxy formation and cosmology.

To measure the galaxy 2PCF, we usually need a random sample with 
the same sky coverage and radial selection function as the galaxy sample
\citep{1993ApJ...417...19H}. For most redshift surveys, the observed
galaxies are flux-limited samples suffering from 
luminosity-dependent selection bias. As the redshift increases, only
luminous galaxies can be observed and the dim galaxies are too faint
to be detected. As a result, the galaxy number density varies as a
function of redshift. Generally, it is easy to produce random samples
for the luminosity-selected galaxies if the luminosity function is
fairly determined. However, for a galaxy sample selected by other
physical quantities such as color, stellar mass, morphology and so
forth, it is not straightforward to generate their corresponding
random samples.

For these property-selected galaxy samples, the shuffled 
method has been widely used in galaxy clustering analysis
\citep{2006MNRAS.368...21L, 2012MNRAS.426.2719R,
  2012MNRAS.427.3435A,2012MNRAS.425..415S,
  2013ApJ...767..122G,2014MNRAS.437.1109R}. Previous tests have shown
that the 2PCF measured using random sample constructed from the
shuffled method produces the least biased result compared with other
methods \citep{2010ApJ...710.1444K,2015MNRAS.449..848H}. Particularly,
\cite{2012MNRAS.424..564R} proved that the systematic bias induced by
the shuffled method is quite small for the redshift-space correlation
function on the scale around $30\sim 150\mpch$, with a statistical
uncertainty of at most $5\%$. However, as current and future redshift
surveys are aiming at $\sim 1\%$ level accuracy of clustering
measurements, the systematic bias induced by the shuffled method
should be carefully taken into account.  Generally, in the shuffled
method, there is a hidden issue that the
structures in the radial distribution of real galaxies can be transferred
to the random sample through the shuffling process, resulting in an
underestimation of galaxy clustering. By applying different approaches
to construct the random samples, \cite{2013A&A...557A..54D} found that
the projected 2PCF (hereafter P2PCF) measured using a random sample from
the \vmax\ method is more accurate than the measure from the shuffled
method. This is not surprising since the random redshifts generated
from the \vmax\ method are randomly distributed in the maximum
observable volume of the galaxies, only depending on the flux limits
of the survey \citep{2011MNRAS.416..739C}. Therefore, in principle,
the \vmax\ method is superior to the shuffled method.

The purpose of this paper is to demonstrate that apart from the
P2PCF, the shuffled method can impact the {\it shape} of the
2PCF in a 2D space, thus result in systematic errors in the redshift-space distortion 
measurement. While such kind of systematics is not induced in the
\vmax\ method. Here we use mock galaxy catalogs to quantify and
compare the systematic uncertainty induced by random samples from the
\vmax\ method and the shuffled method. We primarily focus our 
tests on the galaxy clustering on scales below $40\mpch$.
For the \vmax\ method, we also need to correctly estimate the maximum 
observable volume for individual galaxies based on the magnitude limits 
of the survey.

The paper is organized as follows. In Section~\ref{sect:data}, we
first introduce how we construct the mock galaxy catalogs and prepare
for our tests. Three radial distribution functions that we applied to
produce the random samples are also outlined in this section.  In
Section~\ref{sect:clustering}, we compare the galaxy correlation
functions measured from three different methods in detail and quantify
the systematic uncertainties of these measurements. Finally, we discuss
our results and conclude the paper in Section~\ref{sect:conclusion}.
In our distance calculation, we assume a flat $\Lambda$CDM cosmology
with $\Omega_m=1-\Omega_\Lambda=0.268$,
$h=H_0/(100\,\kms\mathrm{Mpc}^{-1})=0.71$.

\section{Data}
\label{sect:data}

\subsection{Construction of mocks}
\label{sect:mock}

The mock galaxy catalogs are constructed basically in the same way as
\cite{2019ApJ...872...26Y}. Briefly, we use a cosmological $N$-body
simulation from the CosmicGrowth simulation suite
\citep{2019SCPMA..6219511J} named $\rm WMAP\_3072\_600$. This
simulation was performed by executing a parallel adaptive $\rm P^3M$
code with $3072^3$ particles in a $600\mpch$ cube box, assuming a
standard flat $\Lambda \rm CDM $ cosmology with
$\{\Omega_m=0.268,~\Omega_b=0.045, ~\sigma_8=0.83,~n_s=0.968\}$ and
$h=H_0/(100\,\kms\mathrm{Mpc}^{-1})=0.71$, which are consistent with
the observation of the Nine-Year Wilkinson Microwave Anisotropy Probe
(WMAP 9) \citep{2013ApJS..208...20B, 2013ApJS..208...19H}. For each
output snapshot, the friends-of-friends algorithm
\citep{1985ApJ...292..371D} is applied to find halos with a linking
length of 0.2 in units of the mean particle separation. Then, the
Hierarchical Bound-Tracing technique \citep{2012MNRAS.427.2437H,
  2018MNRAS.474..604H} was used to identify subhalos along with their
merger history. We pick the snapshot at $z=0$ to construct our halo
catalog and halos containing at least 50 particles are included.

There are many popular galaxy-halo connection models that can
successfully reproduce the observed galaxy clustering on different
scales. Here, we apply the SHAM model to build the mock galaxy catalogs
by assuming a monotonic relation between the galaxy absolute magnitude
$M^{0.1}_{\rm r}$ and the peak mass $M_{\rm peak}$ of subhalos. The
$M_{\rm peak}$ is defined as the maximum mass that a subhalo ever had
throughout its evolutionary history.  The luminosity function of SDSS DR7
\full\ sample of the New York University Value-Added catalog
(NYU-VAGC)\footnote{$\tt lfvmax-q2.00a-1.00.dr72full1.fits$.}
\citep{2001AJ....121.2358B,2003ApJ...592..819B,2005AJ....129.2562B} is
adopted to perform the SHAM, where the $r-$band absolute magnitude
$M^{0.1}_{\rm r}$ of galaxies have been $k-$ and $e-$corrected to redshift
$z=0.1$. The `orphan' galaxies are also taken into account in the halo
catalog, see \cite{2019ApJ...872...26Y} for details.  A galaxy
naturally obtains the position and velocity of a subhalo when it
matches to the subhalo.  By stacking the simulation box periodically
and randomly setting the locations of the observers, we construct 60
mock galaxy catalogs in total. Galaxies in these mocks are complete at
$M^{0.1}_{\rm r} \le -18$, and their number density $n(z)$ should be the
input $n_{\rm DR7}(z)$ but with a scatter due to cosmic variance. All
mock galaxy catalogs have the same sky coverage of $ \sim 2777 \rm deg^2$
and the same radial comoving distance $d_{\rm C}$ of [0, 600] $\mpch$.
The true redshifts of galaxies are converted into the observed redshift
$z_{\rm obs}$ by adding the influence of peculiar velocity. The
apparent magnitude $m_{\rm r}$ is simply derived by
$m_{\rm r}=M^{0.1}_{\rm r} + 5\mathrm{log}_{10} [d_{\rm C}(1+z_{\rm obs})]+25$.
In this study, as our vital goal is to identify the systematic bias in
clustering measurements caused by different types of random samples,
we use relatively simple models in constructing our mock galaxy
catalogs to eliminate potential uncertainties. First, we do not add a
scatter in the $M^{0.1}_{\rm r}-M_{\rm peak}$ matching relation, so that the
galaxies selected in each realization corresponding to the
same mass subhalos.  Second, we ignore the $k-$ and $e-$corrections in
all magnitude-related calculations. These simplifications will allow us to 
focus on testing the impact of random samples.

\subsection{Mock galaxy catalog}
\label{sect:gsample}

\begin{figure} 
   \centering
   \includegraphics[width=14.0cm, angle=0]{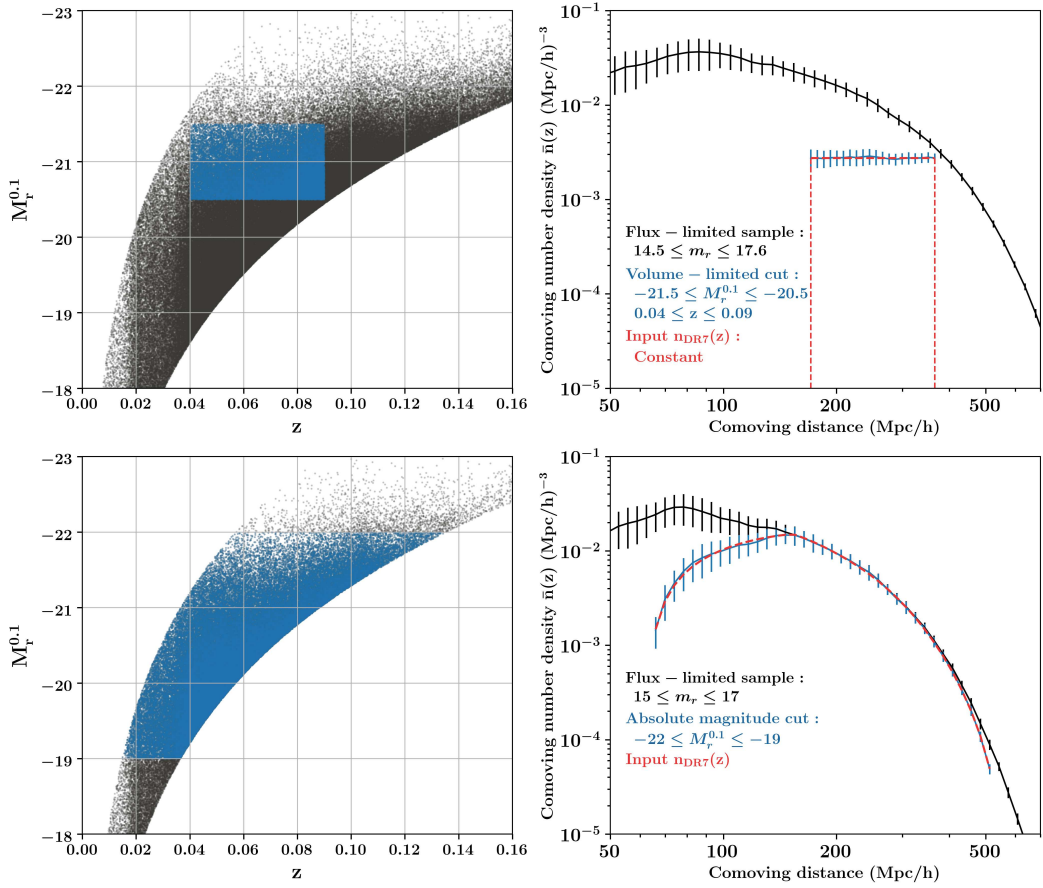}
  \caption{Galaxy distribution in redshift and $r$-band absolute magnitude for the mock galaxy catalog with 
   different selection criteria (left panels), and comparison of the comoving number densities for these 
   samples (right panels). In the upper-left panel, the gray points denote one of our flux-limited samples 
   with apparent magnitude cuts $14.5\le m_{\rm r}\le17.6$. We construct a volume-limited sample from this 
   flux-limited sample as shown in the blue points, that we select galaxies within an absolute magnitude of 
   [$-20.5$, $-21.5$] and a redshift range of [$0.04$, $0.09$]. The black curve in the upper-right panel is the mean 
   comoving number density $\bar{n}(z)$ as a function of comoving distance for the 60 flux-limited mock samples. 
   The error bars denote $1\sigma$ variation among these samples. The $\bar{n}(z)$ of the volume-limited 
   samples are shown in the blue curve with error bars, which is constant as expected. 
   The red dashed line represents the input $n_{\rm DR7}(z)$ of SDSS DR7 \full\ sample. 
   For a flux-limited sample with $15\le m_{\rm r}\le17$ as shown in the lower-left panel (in gray), 
   we further make absolute magnitude cuts at $-22$ and $-19$ (in blue). The mean number density of different 
   flux-limited samples is shown in the lower-right panel. Note, we use $h=H_0/(100\,\kms\mathrm{Mpc}^{-1})=0.71$ 
   in our calculations (see text for details).}
   \label{Fig: gaxsample}
   \end{figure}

Galaxies observed in redshift surveys are usually flux-limited whose
number density may vary as a function of redshift. In order to obtain
a well-understood sample of galaxies for the measurement and modeling
of the two-point statistics, a volume-limited sample or a 
magnitude cut flux-limited sample is usually constructed, at a cost of
discarding a significant number of galaxies, therefore, lower the
statistical accuracy \citep{2005ApJ...630....1Z,2016MNRAS.460.3647X}. 
In this work, we construct both types of galaxy samples to carry out our tests.

Firstly, to define a volume-limited sample, we draw a flux-limited
sample with apparent magnitude $14.5\le m_{\rm r}\le 17.6$ from each mock
sample.  Then, we specify an absolute magnitude range
$-21.5\le M^{0.1}_{\rm r} \le -20.5 $ and a redshift range
$0.04 \le z_{\rm obs}\le 0.09$ to the flux-limited sample, ensuring
that a galaxy in the volume-limited sample can be displaced to any
redshift in [0.04, 0.09] and still remains within the apparent
magnitude limits \citep{2001MNRAS.328...64N, 2002MNRAS.332..827N,
  2004ApJ...606..702T,2011ApJ...736...59Z}.  These constraints result
in a constant comoving number density $n_{\rm const}$, and so is the
radial selection function, hence, it is straightforward to create a
random sample having exactly the same $n_{\rm const}$ as a
volume-limited galaxy sample.

After that, we construct a set of magnitude cut flux-limited samples
with apparent magnitude limits of $15\le m_{\rm r}\le 17$ and absolute
magnitude $-22 \le M^{0.1}_{\rm r} \le -19 $ from the mocks. For a magnitude
cut flux-limited sample, the galaxy number density is a strong
function of redshift $n(z)$, as at a given redshift galaxies only in a
certain absolute magnitude range can be detected by survey
\citep{2002ApJ...571..172Z}. The derivation of galaxy radial selection
function needs to integrate the luminosity function of galaxy sample
appropriately. In our case, we derive the expected comoving number
density as a function of redshift for our flux-limited samples by
equation:
\begin{equation}\label{eq:n(z)}
n(z)=\int^{M^{0.1}_{\rm r,faint}(z)}_{M^{0.1}_{\rm r, bright}(z)} \Phi(M^{0.1}_{\rm r}) dM^{0.1}_{\rm r},
\end{equation}
where $ \Phi(M^{0.1}_{\rm r})$ is the input luminosity function of the SDSS
DR7 \full\ sample, and
\begin{eqnarray}\label{eq:dm}
M^{0.1}_{\rm r, bright}(z)=\mathtt{max}[M^{0.1}_{\rm r, min},15-DM(z)], \\
M^{0.1}_{\rm r, faint}(z)=\mathtt{min}[M^{0.1}_{\rm r, max},17-DM(z)],
\end{eqnarray}
where $DM(z)$ is the distance modulus at redshift $z$, and we set
$M^{0.1}_{\rm r, min}=-22$ and $M^{0.1}_{\rm r, max}=-19$, respectively. The
radial selection function $\phi(z)$ of the flux-limited sample can be
estimated via equation:
\begin{equation}\label{eq:rsf}
\phi(z)=\frac{n(z)}{\int^{-19}_{-22} \Phi(M^{0.1}_{\rm r}) dM^{0.1}_{\rm r}}.
\end{equation}

As an example, we show a volume-limited sample and a flux-limited
sample and the mean number densities of mock galaxy catalogs in
Figure~\ref{Fig: gaxsample}.  In the upper-left panel, the blue points
denote the galaxy distribution of the volume-limited sample on the
redshift and absolute magnitude diagram, the raw flux-limited sample
is denoted by the gray points. In the upper-right panel, the black and
blue curves represent the mean number densities $\bar{n}(z)$ for the
60 flux-limited samples and volume-limited samples, respectively. The
error bars stand for $1\sigma$ variation among these samples. The red
dashed line marks the number density $n_{\rm DR7}(z)$ derived from the
input luminosity function of SDSS DR7 \full\ sample. As expected, the
$\bar{n}(z)$ of volume-limited samples agree very well with the
constant $n_{\rm DR7}(z)$.  The distributions of flux-limited samples
are displayed in the lower panels of Figure~\ref{Fig: gaxsample}. The
mean $\bar{n}(z)$ of the flux-limited sample is a strong function of
redshift, which again agrees with the $n_{\rm DR7}(z)$ estimated from
equation~(\ref{eq:n(z)}) very well.  Once the mean comoving number
density is well estimated, we can easily construct the radial
distribution of random samples for individual galaxy samples.

\subsection{Random sample}
\label{sect:random}

\begin{figure*} 
   \centering
   \includegraphics[width=15.0cm, angle=0]{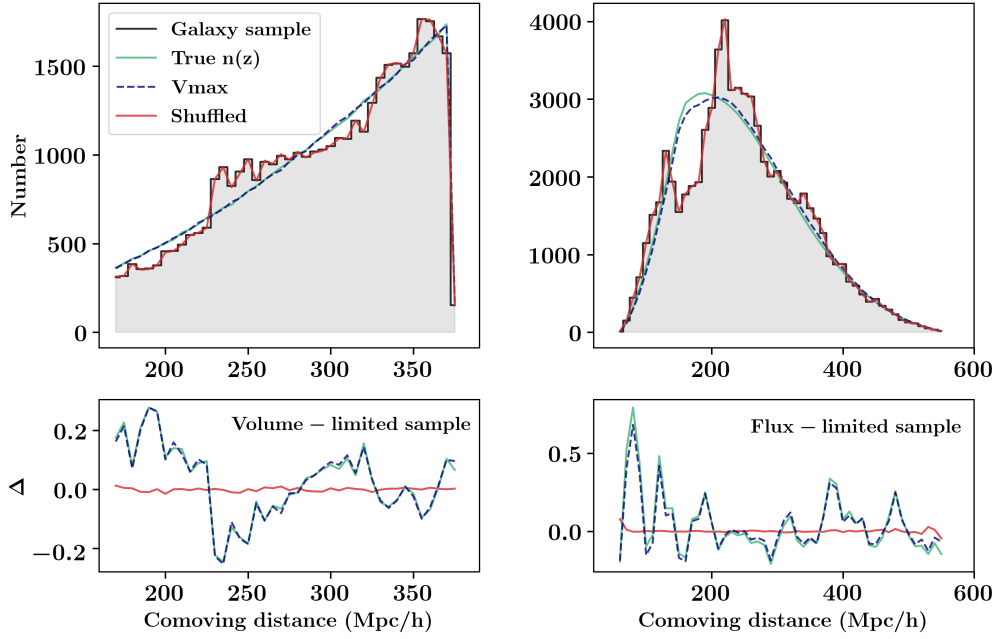}
   \caption{Comparison of radial distributions for galaxy sample and
     random samples. The shaded black histograms denote the
     distributions of a volume-limited sample (left panel) and a
     flux-limited sample (right panel) selected from our 60 mock
     galaxy samples. The bin sizes are $\Delta d=5\mpch$ for
     the volume-limited sample and $\Delta d=10\mpch$ for the flux-limited
     sample, respectively. The green, dashed blue and red curves
     represent the distributions of random samples generated by using
     the radial selection function derived from the true $n(z)$, the
     \vmax\ method, and the shuffled method, separately. The
     total number of each random sample is scaled to be the same
     number as the galaxy sample. The lower small panels show the
     relative number bias in each distance of random points and
     galaxies, which is defined as $\Delta\equiv (n_{\rm r}-n_{\rm g})/n_{\rm g}$. 
     $\Delta$ for different methods are coded in the
     same colors as the upper panels.}
   \label{Fig:hist}
\end{figure*}

\begin{figure} 
   \centering
   \includegraphics[width=10.0cm, angle=0]{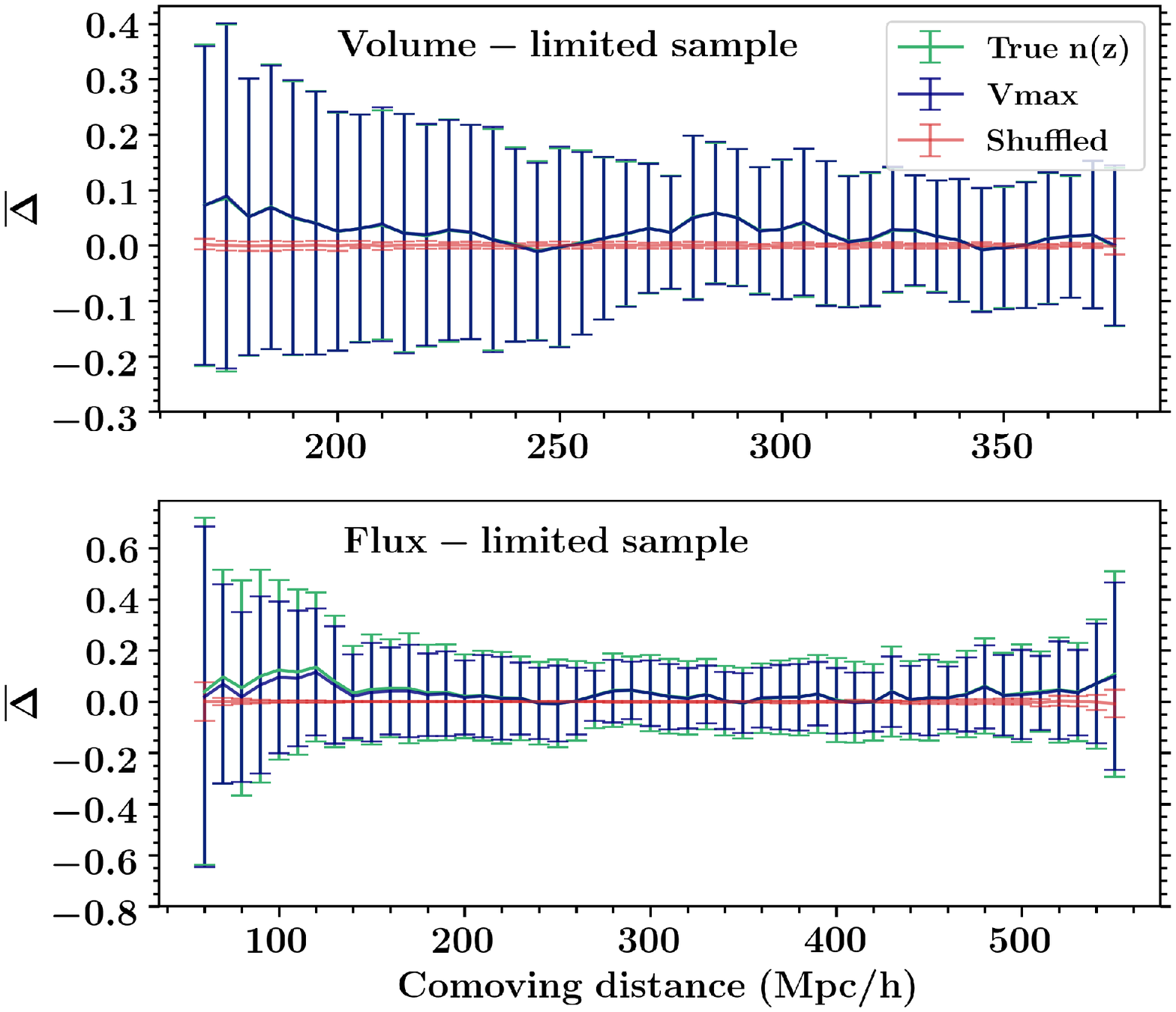}
   \caption{Mean of bias $\overline{\Delta}$ for 60 volume-limited
     samples (upper panel) and 60 flux-limited samples (lower panel),
     respectively. The green, blue and red curves denote the
     $\overline{\Delta}$ for the true $n(z)$, the \vmax\ method, and
     the shuffled method, separately. Error bars stand for $1\sigma$
     deviation of biases among these samples.}
   \label{Fig:meanbias}
\end{figure}

In this study, our basic goal is to identify the systematic
uncertainty in galaxy clustering caused by random samples. More
specifically, we aim to make a robust comparison of the \vmax\ method
and the shuffled method. The comparison will help us assess
to what extent the random samples can impact our measurements of the
2PCFs.

Basically, we construct random samples for individual galaxy samples
based on their radial distributions from three methods. First, we
create a set of random points that uniformly distribute on the surface
of a sphere, then points covering an equal area as the galaxy sample
is selected.  Without adding any angular selection effect, we take
these points as the angular positions of the random samples.
Theoretically, modeling the radial distribution of galaxy sample
requires the true number density of galaxy sample. This is difficult
to achieve in observation since we always sample galaxies in a certain
volume of the universe, and the $n(z)$ can only be estimated
empirically from the observed galaxies. By using mocks, the true
number density $n(z)$ is the input $n_{\rm DR7}(z)$ as described in
Section~\ref{sect:gsample}, therefore, allowing us to construct the
$n(z)$ for random samples exactly identical to the true one. Three
methods that are used to construct the radial distribution are
described below:
\begin{enumerate}
\item \textbf{True} $n(z)$, where we apply the radial selection
  function $\phi(z)$ derived from equation~(\ref{eq:n(z)}) with the
  input $n_{\rm DR7}(z)$ to build the redshifts.  In the following
  tests, we will use the correlation functions measured using the true
  $n(z)$ as the benchmarks and explore impacts of the \vmax\ method or
  the shuffled method.

\item \textbf{V}$_{\mathbf{max}}$ method, where we uniformly spread
  random points in the maximum observable volume \vmax\ of individual
  galaxies to obtain the radial comoving distance for random
  samples. For a volume-limited sample, \vmax\ is a fixed volume at
  redshift range [0.04, 0.09]. For a flux-limited sample, we assign
  the absolute magnitudes of observed galaxies to random points, and
  their maximum/minimum redshifts $z_{\rm max, min}$ are estimated by
  \begin{eqnarray}
   z_{\rm max}=\mathtt{min}[z_{\rm m, max},z_{\rm sample, max}],\\
   z_{\rm min}=\mathtt{max}[z_{\rm m, min},z_{\rm sample, min}].
  \end{eqnarray}\label{eq:zz2}
  and 
  \begin{eqnarray}
  m_{\rm r,faint}=M^{0.1}_{\rm r,faint}+DM(z_{\rm m, max}),\\
  m_{\rm r,bright}=M^{0.1}_{\rm r,bright}+DM(z_{\rm m, min}).
  \end{eqnarray}\label{eq:mm2}
  where we set $m_{\rm r,bright}=15$,  $m_{\rm r,faint}=17$, $M^{0.1}_{\rm r,bright}=-22$, 
  and $M^{0.1}_{\rm r,faint}=-19$ based on the magnitude cuts to the
  flux-limited samples.

\item \textbf{Shuffled} method, where we randomly select
  redshifts from the galaxy samples and assign the redshifts to the
  corresponding random samples.
\end{enumerate}

As an example, Figure~\ref{Fig:hist} demonstrates the distribution of
radial comoving distance for the two types of galaxy samples (shaded
black histograms) in one of our realizations, as well as the random
samples generated using the radial selection function $\phi(z)$
estimated from the true $n(z)$ (green curves), the \vmax\ method (blue
dashed curves), and the shuffled method (red curves),
respectively.  The bin size is $\Delta d=5\mpch$ for the
volume-limited sample and $\Delta d=10\mpch$ for the flux-limited
sample. We also compute the number difference $\Delta$ of random
samples relative to the galaxy samples as shown in the lower small
panels. We define the difference as $\Delta\equiv (n_{\rm r}-n_{\rm g})/n_{\rm g}$,
where $n_{\rm r}$ and $n_{\rm g}$ denote the number of random points
and galaxies in each distance bin. The mean difference
$\overline{\Delta}$ of individual $\Delta$ for all 60 realizations are
displayed in Figure~\ref{Fig:meanbias}.  The error bars represent the
standard deviation from $\overline{\Delta}$ among all samples in each
bin. Apparently, random samples constructed using the shuffled method
show the best agreement with the radial distribution of galaxies for
both the volume-limited samples and the flux-limited samples,
indicating the structures of galaxies in the line-of-sight direction
are reserved.  Radial distributions constructed by the true $n(z)$ and
the \vmax\ method are nearly identical, especially for the
volume-limited samples. Meanwhile, the two distributions exhibit small
deviations from each other in the flux-limited sample.  As mentioned
before, this is due to the fact that each observation is actually one
sampling of a small set of galaxies in the universe. The larger the
observed galaxy sample, the closer the number density is to the true
$n(z)$. Moreover, we see that the distribution given by the \vmax\
method seems slightly closer to the distribution of the galaxy
sample. This difference indicates that the \vmax\ method still suffers
very slightly from the large-scale structure as noted by
\cite{2005ApJ...631..208B}, which may impact the luminosity function
and hence the redshift distribution of our random points.

\section{Clustering measurement}
\label{sect:clustering}

In this section, we will compare galaxy correlation functions measured
using three different random samples, to demonstrate that using the
one constructed from the shuffled method leads to an
underestimation of galaxy clustering. While the measurements using the
\vmax\ random sample has much better performance on all
scales that we explored.

\subsection{Clustering estimator}
\label{sect:estimator}

We use the common way to calculate the correlation function
\citep{1988ASPC....4..257H,1992ApJ...385L...5H, 1994MNRAS.266...50F}
in a 2D space, that the redshift separation vector $\mathbf{s}$ and
the line-of-sight vector $\mathbf{l}$ are defined as
$\mathbf{s} \equiv \mathbf{v}_1-\mathbf{v}_2 $ and
$\mathbf{l}\equiv (\mathbf{v}_1+\mathbf{v}_2)/2$, separately, where
$\mathbf{v}_1$ and $\mathbf{v}_2$ are the redshift space position
vectors of a pair of galaxies. The separations parallel ($\pi$) and
perpendicular ($r_p$) to the line of sight are derived as
\begin{equation}
 \pi \equiv \frac{\mathbf{s}\cdot \mathbf{l}}{|\mathbf{l}|}, ~~~~~~r^2_p \equiv \mathbf{s}\cdot \mathbf{s}-\pi^2.
\end{equation}
 A grid of $\pi$ and $r_p$ is
constructed by taking $1\mpch$ as the bin size for $\pi$ from 0 linearly up to $\pi_{\rm max}=60\mpch$ 
and $0.2$ dex as the bin size for $r_p$ logarithmically in the range of [$0.01$, $60$] $\mpch$. 
The estimator of \cite{1993ApJ...412...64L} is adopted as
\begin{equation}
 \xi(r_p,\pi)  = \frac{DD-2DR+RR}{RR},
\end{equation}
where $DD$, $DR$, and $RR$ are the numbers of data-data, data-random,
and random-random pairs. Then, by integrating the $\xi(r_p,\pi)$ along the line-of-sight separation we
estimate the P2PCF \citep{1983ApJ...267..465D} by 
\begin{equation}
w_p(r_p)\equiv 2\int^{\infty}_0 \xi(r_p,\pi)~d\pi = 2\int^{\pi_{max}}_0 \xi(r_p,\pi)~d\pi.
\end{equation}
In this work, we run $\tt{CORRFUNC}$ \citep{10.1007/978-981-13-7729-7_1} for pair counting to measure 
all mock galaxy correlation functions. In order to reduce the shot noise, we use random samples which 
are $\sim$ 40 times the number of galaxies.

\subsection{Comparison of correlation functions}
\label{sect:cf}

\begin{figure*} 
   \centering
   \includegraphics[width=12.0cm, angle=0]{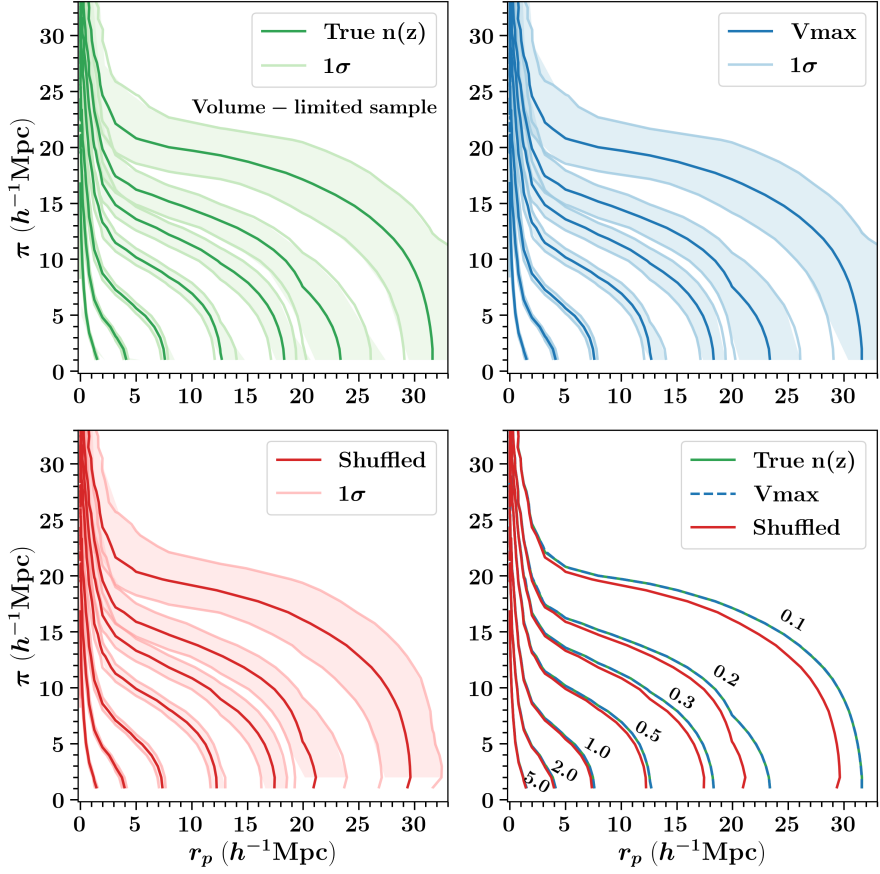}
   \caption{The average correlation function contours $\overline{\xi}(r_p,\pi)$ measured using random 
   samples constructed with the true $n(z)$ (top-left panel), the \vmax\ method (top-right panel), 
   and the shuffled method (bottom-left panel) for the volume-limited samples, respectively. 
   The shaded regions mark $1\sigma$ deviations among 60 mock samples. The bottom-right 
   figure shows comparison of $\overline{\xi}(r_p,\pi)$ for different methods. The contour levels from 
   outside-in correspond to $\overline{\xi}(r_p,\pi)=[0.1,0.2,0.3,0.5,1,2,5]$, respectively.} 
   \label{Fig:vcontours}
\end{figure*}

\begin{figure*} 
   \centering
   \includegraphics[width=12.0cm, angle=0]{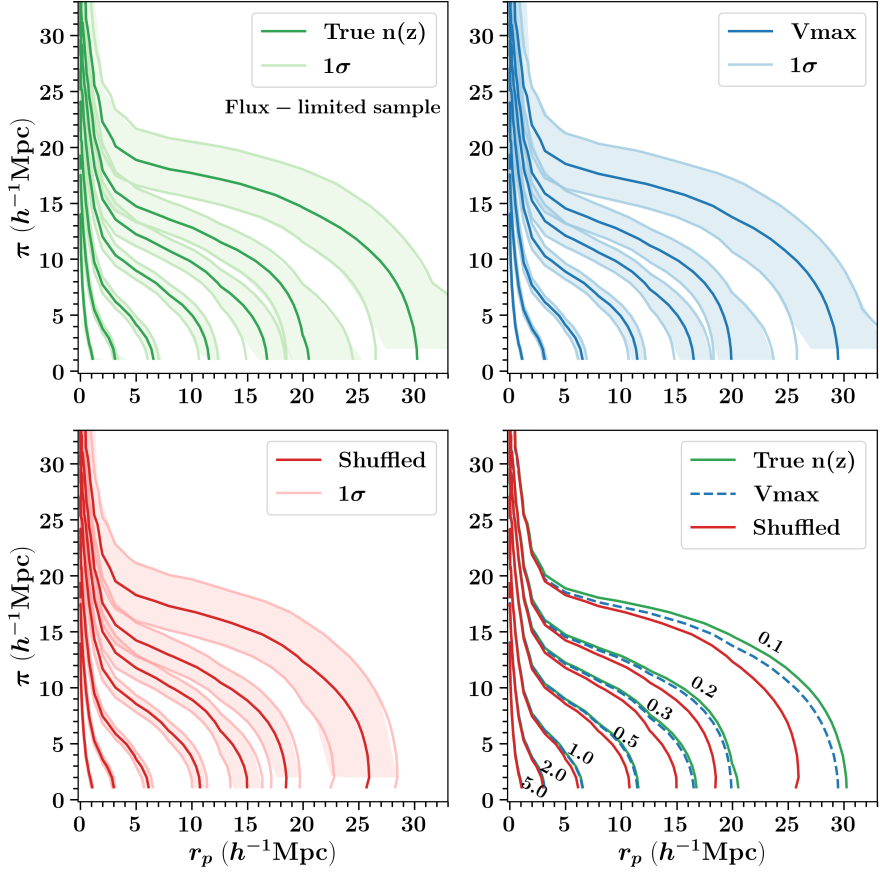}
   \caption{The same as Figure~\ref{Fig:vcontours} but for the flux-limited samples.} 
   \label{Fig:fcontours}
\end{figure*}

\begin{figure*} 
   \centering
   \includegraphics[width=15.0cm, angle=0]{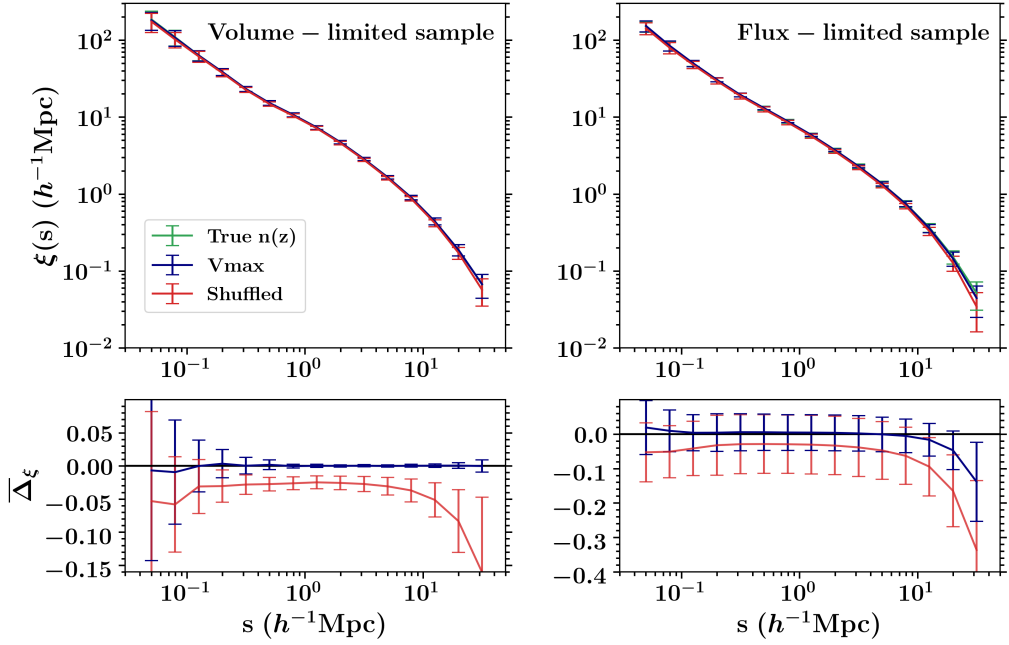}
   \caption{Top panels: The average redshift-space correlation functions $\xi(s)$ of 60 volume-limited samples 
   (left) and 60 flux-limited samples (right) and their standard deviations (error bars). The green curves represent 
   the true $\xi(s)$ measured using random samples from the true $n(z)$, the blue and red curves denote the 
   $\xi(s)$ for the \vmax\ method and the shuffled method, respectively. Bottom panels: The average bias 
   $\overline{\Delta}_{\xi}$ and $1\sigma$ deviations from the true $\xi(s)$ for the two radial selection models, 
   determined using 60 mock galaxy catalogs. } 
   \label{Fig:xis}
\end{figure*}

\begin{figure*} 
   \centering
   \includegraphics[width=15.0cm, angle=0]{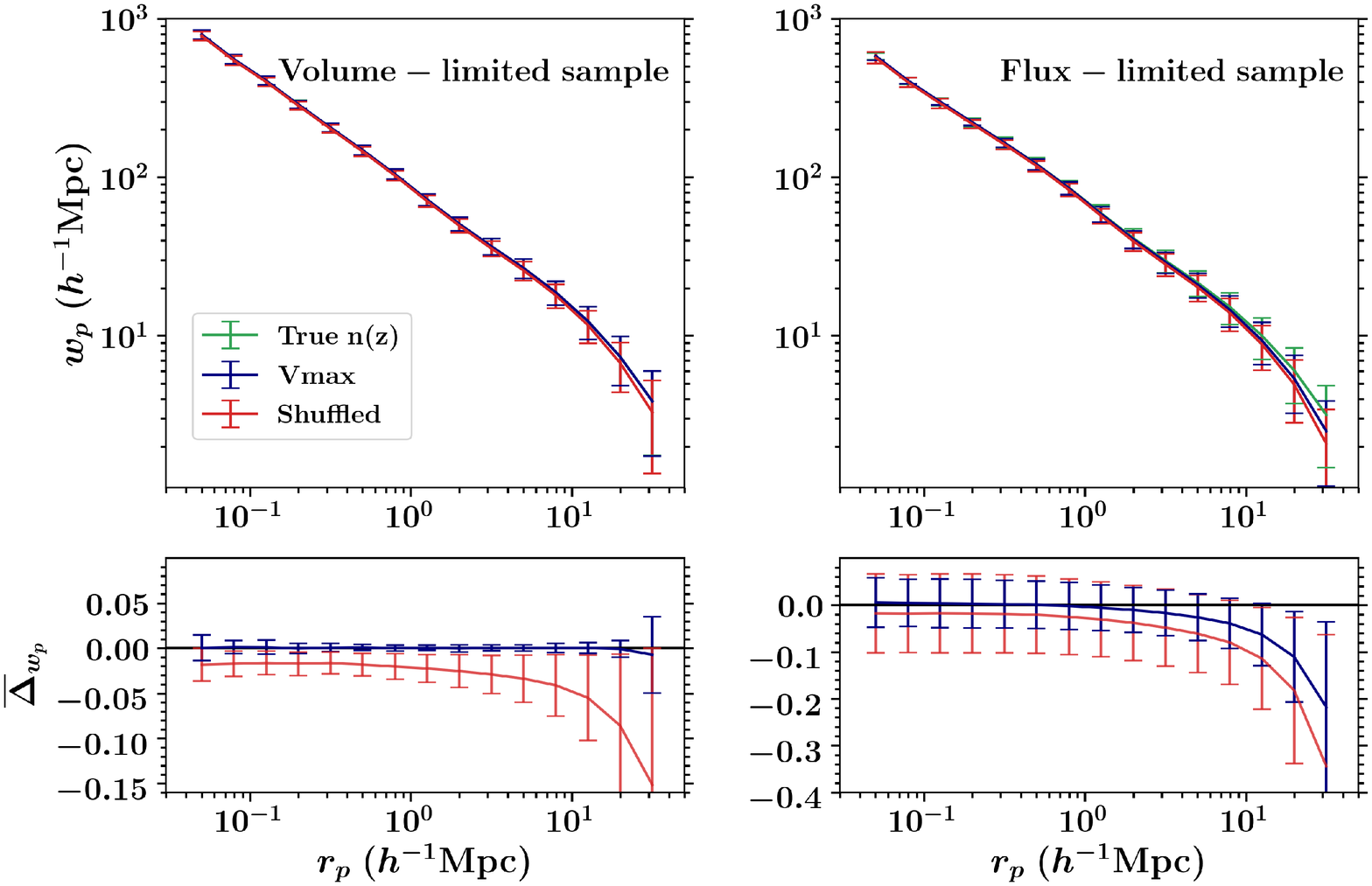}
   \caption{The same as Figure~\ref{Fig:xis} but for the projected two-point correlation functions.} 
   \label{Fig:wps}
\end{figure*}

Our main results in comparison of the 2PCFs measured from three
different methods are presented in this section. We consider the
correlation function measured from the true $n(z)$ as the true 2PCF.
The comparison of the correlation function contours, the
redshift-space correlation functions, and the projected two-point
correlation functions are shown in Figure~\ref{Fig:vcontours} to
Figure~\ref{Fig:wps}, respectively.

Figure~\ref{Fig:vcontours} and Figure~\ref{Fig:fcontours} display the
average contours of the two-dimensional correlation functions
$\xi(r_p,\pi)$ for the volume-limited samples and the flux-limited
samples, respectively. The true $\xi(r_p,\pi)$ (hereafter
$\xi_{\rm true}(r_p,\pi)$) derived from the true $n(z)$ is denoted by
the green contours in the top-left panel, the shaded light green
regions represent $1\sigma$ variance among 60 individual measurements
of $\xi_{\rm true}(r_p,\pi)$. The blue and red contours in the
top-right panel and the bottom-left panel denote the $\xi(r_p,\pi)$
of the \vmax\ method and the shuffled method (hereafter
$\xi_{\rm Vmax}(r_p,\pi)$ and $\xi_{\rm shuffled}(r_p,\pi)$) , 
separately. Comparison of the average $\xi(r_p,\pi)$ for all three
different methods is shown in the bottom-right panel, where
$\xi_{\rm Vmax}$ is denoted by the dashed blue contours to distinguish
from $\xi_{\rm true}$. For the volume-limited samples, the
$\xi_{\rm Vmax}(r_p,\pi)$ contours are generally indistinguishable
from the $\xi_{\rm true}(r_p,\pi)$ contours. For the flux-limited
samples, $\xi_{\rm Vmax}(r_p,\pi)$ contours exhibit overall great
agreement with the true ones, with a very small systematic bias at
large radii well below the $1\sigma$ uncertainty.

For the case of shuffled method, the $\xi_{\rm shuffled}(r_p,\pi)$
exhibit a prominent systematic bias from $\xi_{\rm true}(r_p,\pi)$ for
both types of samples, especially for the flux-limited samples where
the bias is almost beyond $1\sigma$ statistical uncertainty. Note
  that since the systematic bias indeed changes the shape of
  $\xi(r_p,\pi)$, which will induce systematic errors in the
  cosmological probes using the redshift distortion effects on
  intermediate scales \citep[see e.g.][]{2018ApJ...861..137S}.

The comparison of the average redshift-space correlation functions
$\xi(s)$ is shown in the upper panels of Figure~\ref{Fig:xis}. The
left panel displays the mean $\overline{\Delta}_{\xi}$ of 60
volumed-limited mock samples, the right panel shows the same results
but for the flux-limited samples. The true $\xi_{\rm true}(s)$ is
denoted by the green curve. The $\xi(s)$ from the \vmax\ method and
the shuffled method are in blue and red curves,
respectively. The error bars represent $1\sigma$ variance among
$\xi(s)$. The lower panels show the mean bias
$\overline{\Delta}_{\xi}$ of $\xi _{\rm Vmax}(s)$ and
$\xi _{\rm shuffled}(s)$ with respect to $\xi_{\rm true}(s)$. The mean
bias is defined as
$\overline{\Delta}_{\xi}=(\sum^N_{i=1}\Delta^i_{\xi})/N$, where
$\Delta^i_{\xi}=(\xi^i_{\rm Vmax}-\xi^i_{\rm true})/\xi^i_{\rm true}$
for \vmax\ method or
$\Delta^i_{\xi}=(\xi^i_{\rm shuffled}-\xi^i_{\rm true})/\xi^i_{\rm
  true}$ for shuffled method, the $\xi^i$ is the correlation function
of the $i$th galaxy sample, and $N=60$.  The error bars denote
$1\sigma$ variance of 60 individual $\Delta^i_{\xi}$.  We can clearly
see that, the comparison results are completely consistent with the
results shown in Figure~\ref{Fig:vcontours} and
Figure~\ref{Fig:fcontours}.  For the volume-limited samples, the mean
bias of $\xi_{\rm Vmax}$ relative to $\xi_{\rm true}$ is almost
zero. Comparatively, there is a systematic bias of $\sim 3\%$ between
$\xi_{\rm shuffled}$ and $\xi_{\rm true}$ at small scales. On scales 
above $8 \mpch$, the bias gradually increases. At the scale of
$\sim 30 \mpch$, the mean bias $\overline{\Delta}_{\xi, \rm shuffled}$
is up $\sim 15\%$.  For the flux-limited samples, the \vmax\ method
also exhibits much better performance than the shuffled method. On
scale below $10\mpch$, $\xi_{\rm Vmax}$ is fairly identical to
$\xi_{\rm true}$, and $\xi_{\rm shuffled}$ exhibits an underestimate
with a bias up to $\sim 5\%$. On scales $s > 10\mpch$, both methods
display underestimates to a certain extent, where the
$\overline{\Delta}_{\xi, \rm Vmax}$ gradually increases to $12\%$ at
the scale of $30\mpch$ and the $\overline{\Delta}_{\xi, \rm shuffled}$
is $\sim 30\%$ on the same scale.

Finally, comparison of the average P2PCFs are shown in
Figure~\ref{Fig:wps}, where the color-coded $w_p$ from three different
methods are the same as Figure~\ref{Fig:xis}. We can see that, without
the effect of redshift distortion, $w_{p, \rm Vmax}$ remains roughly
identical to $w_{p, \rm true}$ for the volume-limited samples (left
panels), and for the flux-limited samples (right panels)
$w_{p, \rm Vmax}$ also agrees with $w_{p, \rm true}$ on scale smaller
than $1\mpch$. While, on larger scale, the \vmax\ method results in a
slightly increasing underestimation compared with the true one, and
$\overline{\Delta}_{w_{p,\rm Vmax}}$ is also kind of larger than
$\overline{\Delta}_{\xi,\rm Vmax}$. As for the shuffled method, on
scale of $r_p < 1\mpch$, the average $w_{p, \rm shuffled}$ shows less
deviation from $w_{p, \rm true}$ compared with the results of
redshift-space correlation functions for both types of samples. 
The mean deviation of $w_{p, \rm shuffled}$ from $w_{p, \rm true}$ 
increases with a strong bias above scales of $1\mpch$. 
The $\overline{\Delta}_{w_{p,\rm shuffled}}$ is up to $15\%$
for the volume-limited samples and $34\%$ for the flux-limited samples
at $30\mpch$, respectively.

Based on the above comparison, our results steadily demonstrate that using
random samples constructed from the \vmax\ method to measure the
correlation function, we can achieve much higher accuracy than those
from the shuffled method.  For the volume-limited samples, the average
correlation functions from the \vmax\ method are almost identical
to the true correlation functions on all concerned scales, with nearly
zero bias and negligible statistical uncertainty.  For the
flux-limited samples, on scales smaller than $10\mpch$ for $\xi(s)$ and 
$1\mpch$ for $w_p(r_p)$, the average correlation functions from the 
\vmax\ method are still fully consistent with the true correlation functions 
with a statistical uncertainty of at most $5\%$. As the scale increases, 
the correlation functions from the \vmax\ method tends to underestimate 
the galaxy clustering slightly.  Nevertheless, the mean bias from 
the \vmax\ method is still much smaller than the bias from the shuffled 
method in general.

\section{Discussion ans conclusion}
\label{sect:conclusion}

In this paper, using mock galaxy catalogs, we have investigated the
systematic bias induced by random samples generated using the \vmax\
method and the shuffled method in galaxy clustering
measurement. We have compared the redshift-space correlation functions
and the projected 2PCFs for the volume-limited samples and the
flux-limited samples, respectively. Our results demonstrate that the
\vmax\ method is more robust to simulate the radial distribution of
galaxies for the random sample.  Our main results can be summarized as
follows:
\begin{enumerate}
\item For the volume-limited samples, the \vmax\ method can produce an
  unbiased measure of galaxy clustering on the scale less than $40\mpch$,
  while, the shuffled method results in an increasing systematic
  underestimation with the increase of scale.

\item For the flux-limited samples, the 2PCFs measured from random 
samples of the \vmax\ method remain unbiased concerning the true galaxy 
clustering on small scales. While on scales larger than $10\mpch$, 
both methods display a systematic bias beyond the systematic uncertainty, 
but the \vmax\ method still has better performance than the shuffled method.

\item By comparing the correlation contours, we find that the shuffled
  method can significantly underestimate the squashing effect on large
  scales, which may induce potential systematics in cosmological
  probes using the linear redshift distortion effect.

\item Finally, the projected 2PCF measured from the shuffled method
  still produces an underestimation, especially on scales larger than
  $2\mpch$. This scale is also known as the ``two-halo term'' scale
  \citep{1996MNRAS.282..347M,
    1999MNRAS.308..119S,2002PhR...372....1C}. Thus, if galaxy
  clustering measured from the redshift shuffled random samples are
  used as constraints, a non-negligible systematic bias will be
  introduced to models such as the halo model, galaxy formation
  models, and the galaxy-halo connection models.
\end{enumerate}

Based on the above tests, we suggest using the \vmax\ method to generate
random samples. The galaxy correlation function from the \vmax\ method
can recover the galaxy clustering more accurately, then providing more
reliable and stringent constraints on the models of galaxy formation and
cosmology.

Besides, there are some simplifications in our probes to be noted as
well.  In this paper, we ignored the $k-$ and $e-$
corrections in our tests, however, these corrections need to be
carefully handled when the analysis is performed to the observed
galaxies. To determine these corrections, one should fit the spectral 
templates to the galaxy spectrum or the broad-band photometry
\citep{2007AJ....133..734B}, and the fitting results largely depend
on the assumptions of the galaxy star formation history, the stellar
population synthesis model, and the dust extinction model
\citep{2001MNRAS.322..231K, 2012MNRAS.422.3285P}. 
While, as long as the maximum observable volume of individual galaxies 
is estimated correctly, our conclusions still firmly hold.
In addition, we also note that the systematic bias from the shuffled method 
determined by \cite{2012MNRAS.424..564R} is somewhat smaller than 
ours for the redshift-space correlation function at scale $\sim 30 \mpch$. 
One possible reason is that our tests performed with the low redshift 
galaxies of SDSS DR7, with a median redshift at $\sim 0.1$. 
But the BOSS CMASS data that they studied is a high-redshift sample 
with a median redshift at $\sim 0.52$, where they have a larger volume. 
All in all, we are confident in our tests, that the \vmax\
method is a more robust way to measure the galaxy clustering.  We will
adopt this method to investigate the property-dependent galaxy
clustering in our future works.

\normalem
\begin{acknowledgements}
We are grateful to the referee for the detailed review of our paper.
The work is supported by the 973 Program (Nos. 2015CB857002,
2015CB857003) and NSFC (11533006, 11621303, 11833005,
11890691, 11890692).

\end{acknowledgements}
  
\bibliographystyle{raa}
\bibliography{ms20190333}

\end{document}